\documentclass[apjl]{emulateapj}

\usepackage{graphicx}
\usepackage{ifpdf}
\usepackage{ifthen}

\newboolean{emulateapj}
\setboolean{emulateapj}{true}

\newcommand{\ctbd}[1]{}

\newcommand{\lc}{light curve}
\newcommand{\lcs}{light curves}

\newcommand{\cfa}{Harvard-Smithsonian Center for Astrophysics (CfA)}

\newcommand{\kms}{\ensuremath{\rm km\,s^{-1}}}
\newcommand{\ms}{\ensuremath{\rm m\,s^{-1}}}

\newcommand{\gcmc}{\ensuremath{\rm g\,cm^{-3}}}

\newcommand{\teff}{\ensuremath{T_{\rm eff}}}
\newcommand{\logg}{\ensuremath{\log{g}}}
\newcommand{\vsini}{\ensuremath{v \sin{i}}}
\newcommand{\feh}{[Fe/H]}

\newcommand{\rsun}{\ensuremath{R_\sun}}
\newcommand{\msun}{\ensuremath{M_\sun}}
\newcommand{\lsun}{\ensuremath{L_\sun}}

\newcommand{\rstar}{\ensuremath{R_\star}}
\newcommand{\mstar}{\ensuremath{M_\star}}
\newcommand{\lstar}{\ensuremath{L_\star}}

\newcommand{\rpl}{\ensuremath{R_{p}}}
\newcommand{\mpl}{\ensuremath{M_{p}}}

\newcommand{\rhopl}{\ensuremath{\rho_{p}}}
\newcommand{\ipl}{\ensuremath{i_{p}}}

\newcommand{\gpl}{\ensuremath{g_{p}}}

\newcommand{\rjup}{\ensuremath{R_{\rm J}}}
\newcommand{\mjup}{\ensuremath{M_{\rm J}}}

\newcommand{\rjuplong}{\ensuremath{R_{\rm Jup}}}
\newcommand{\mjuplong}{\ensuremath{M_{\rm Jup}}}

\newcommand{\figr}[1]{Fig.~\ref{fig:#1}}
\newcommand{\secr}[1]{\mbox{\S\ \ref{sec:#1}}}

\newcommand{\tabr}[1]{\mbox{Table~\ref{tab:#1}}}

\newcommand{\flwof}{\mbox{FLWO 1.2 m}}

\newcommand{\wom}{\mbox{Wise 1 m}}

\defcitealias{kovacs05}{KBN05}
\defcitealias{fortney06}{FMB06}

\newcommand{\hatcur}{HAT-P-5}
\newcommand{\hatcurb}{HAT-P-5b}

\newcommand{\hatcurm}{\ensuremath{1.06\pm0.11}}
\newcommand{\hatcurr}{\ensuremath{1.26\pm0.05}}
\newcommand{\hatcurrho}{\ensuremath{0.66\pm0.11}}
\newcommand{\hatcuri}{\ensuremath{86\fdg75\pm0\fdg44}}
\newcommand{\hatcurg}{\ensuremath{16.5\pm1.9}}
\newcommand{\hatcurar}{\ensuremath{7.50\pm0.19}}
\newcommand{\hatcurarel}{\ensuremath{0.04075\pm0.00076}}
\newcommand{\hatcurP}{\ensuremath{2.788491\pm0.000025}}
\newcommand{\hatcurT}{\ensuremath{2,\!454,\!241.77663\pm0.00022}}
\newcommand{\hatcurdur}{\ensuremath{0.1217\pm0.0012}}
\newcommand{\hatcuringdur}{\ensuremath{0.0145\pm 0.0007}}

\newcommand{\hatcurrprstar}{\ensuremath{0.1106\pm0.0006}}

\shorttitle{\hatcurb: A Jupiter-like transiting hot Jupiter}
\shortauthors{Bakos et al.}

\begin{document}
\ifthenelse{\boolean{emulateapj}}{
\title{\hatcurb: A Jupiter-like hot Jupiter
	Transiting a Bright Star\altaffilmark{$\dagger$}}}
{\title{\hatcurb: A Jupiter-like hot Jupiter
	Transiting a Bright Star\altaffilmark{\dagger}}}
\author{
	G.~\'A.~Bakos\altaffilmark{1,2},
	A.~Shporer\altaffilmark{3},
	A.~P\'al\altaffilmark{4,1},
	G.~Torres\altaffilmark{1},
	G\'eza Kov\'acs\altaffilmark{5},
	D.~W.~Latham\altaffilmark{1},
	T.~Mazeh\altaffilmark{3},
	A.~Ofir\altaffilmark{3},
	R.~W.~Noyes\altaffilmark{1},
	D.~D.~Sasselov\altaffilmark{1},
	F.~Bouchy\altaffilmark{7},
	F.~Pont\altaffilmark{6},
	D.~Queloz\altaffilmark{6},
	S.~Udry\altaffilmark{6},
	G.~Esquerdo\altaffilmark{1},
	B.~Sip\H{o}cz\altaffilmark{4,1},
	G\'abor Kov\'acs\altaffilmark{1},
	J.~L\'az\'ar\altaffilmark{8},
	I.~Papp\altaffilmark{8} \&
	P.~S\'ari\altaffilmark{8}
}

\altaffiltext{1}{\cfa,
	60 Garden Street, Cambridge, MA 02138, USA; gbakos@cfa.harvard.edu.}
\altaffiltext{2}{Hubble Fellow.}
\altaffiltext{3}{Wise Observatory, Tel Aviv University, Tel Aviv,
    Israel 69978}
\altaffiltext{4}{Department of Astronomy,
	E\"otv\"os Lor\'and University, Pf.~32, H-1518 Budapest, Hungary.}
\altaffiltext{5}{Konkoly Observatory, Budapest, P.O.~Box 67, H-1125, Hungary}
\altaffiltext{6}{Observatoire Astronomique de l'Université de Genève,
	51 chemin des Maillettes, CH-1290 Sauverny, Switzerland}
\altaffiltext{7}{Institut d'Astrophysique de Paris, 98bis Bd Arago,
	75014 Paris, France}
\altaffiltext{8}{Hungarian Astronomical Association, 1461 Budapest, 
	P.~O.~Box 219, Hungary}
\altaffiltext{$\dagger$}{
	Based in part on radial velocities obtained with the SOPHIE
	spectrograph mounted on the 1.93m telescope at
	the Observatory of Haute Provance (run 07A.PNP.MAZE).
}
\setcounter{footnote}{1}

\begin{abstract}
	We report the discovery of a planet transiting a moderately bright
	($V=12.00$) G star, with an orbital period of $2.788491\pm0.000025$
	days. From the transit light curve we determine that the radius of
	the planet is $\rpl = 1.257\pm0.053\,\rjuplong$. \hatcurb\ has a
	mass of $\mpl = \hatcurm\,\mjuplong$, similar to the average mass
	of previously-known transiting exoplanets, and a density of $\rhopl
	= \hatcurrho\,\gcmc$. We find that the center of transit is
	$T_{\mathrm{c}} = \hatcurT$ (HJD), and the total transit duration
	is \hatcurdur\,days.
\end{abstract}

\keywords{
	stars: individual: {\mbox GSC 02634-01087} \---
	planetary systems: individual: \hatcurb
}

\section{Introduction}
\label{sec:intro}

To date about 20 extrasolar planets have been found which transit their
parent stars and thus yield values for their mass and
radius\footnote{Extrasolar Planets Encyclopedia: http://exoplanet.eu}.
Masses range from 0.07\,\mjup\ \citep[GJ436;][]{gillon07} to about
9\,\mjup\ \citep[HAT-P-2b;][]{bakos07}, and radii from 0.7\,\rjup\
(GJ436) to about 1.7\,\rjup\ \citep[TRES-4;][]{mandushev07}. These data
provide an opportunity to compare observations with theoretical models
of planetary structure across a wide range of parameters, including
those of the host star \citep[e.g.][and references
therein]{burrows07,fortney07}. Transits also yield precise
determination of other physical parameters of the extrasolar planets,
for instance the surface gravity. Interesting correlations between
these parameters have been noted early on, such as that between masses
and periods \citep{mazeh05} or periods and surface gravities
\citep{southworth07}. Classes of these close-in planets have also been
suggested, such as very hot Jupiters (VHJs; P=1--3 days) and hot
Jupiters \citep[HJs; P=3--9 days;][]{gaudi05}, or a possible dichotomy
based on Safronov numbers \citep{hansen07}. However, the small ensemble
of transiting exoplanets (TEPs) does not allow robust conclusions, thus
the addition of new discoveries is valuable.

Over the past year the HATNet project\footnote{www.hatnet.hu}
\citep{bakos02,bakos04}, a wide-angle photometric survey, has announced four
TEPs. In this Letter we report on the detection of a new transiting
exoplanet, which we label \hatcurb, and our determination of its
parameters, such as mass, radius, density and surface gravity.

\section{Observations and Analysis}
\label{sec:anal}

\subsection{Detection of the transit in the HATNet data}
\label{sec:dete}

\notetoeditor{This is the intended place of \figr{lc}. We would like to 
typeset it as a single column figure.}

\ifthenelse{\boolean{emulateapj}}{\begin{figure}[t]}{\begin{figure}[t]}
\ifpdf
\plotone{img/lc.pdf}
\else
\plotone{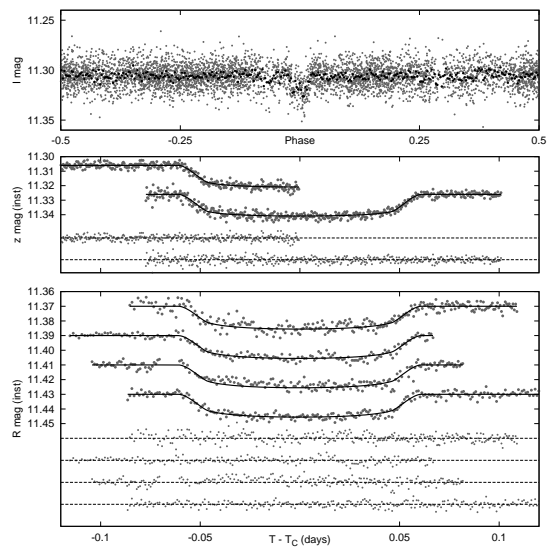}
\fi
\caption{
	The top panel shows the unbinned HATNet \lc\ with 4940 data points,
	phased with the period $P=2.788491$\,d, and with the binned data
	overplotted. The 0.013\,mag deep transit is detected with a dip
	significance of 18. The other panels show photometry follow-up in
	the following order: Sloan $z$-band photometry taken with the
	\flwof\ telescope (on two separate dates: UT 2007 May 18 and UT
	2007 May 21), Cousins $R$-band photometry taken with the \wom\
	telescope (on 4 nights: UT 2007 May 26, June 20, July 4 and 18). 
	Over-plotted are our best analytic fits as described in the text.
\label{fig:lc}}
\ifthenelse{\boolean{emulateapj}}{\end{figure}}{\end{figure}}

GSC~02634-01087, also known as 2MASS J18173731+3637170 is a G star with
$I\approx11.3$ and $V\approx12.00$. It was initially identified as a
transit candidate in our internally labeled field G196, centered at
$\alpha=18^{\mathrm{h}}08^{\mathrm{h}}$, $\delta = 37\arcdeg 30'$. The
data were acquired by HATNet's HAT-7 telescope at the Fred Lawrence
Whipple Observatory (FLWO) of the Smithsonian Astrophysical Observatory
(SAO) and HAT-9 telescope at the Submillimeter Array (SMA) site atop
Mauna Kea, Hawaii. Following the standard calibration procedure of the
frames, data were reduced using the astrometry code of \citet{pal06},
and a highly fine-tuned aperture photometry. We applied our external
parameter decorrelation (EPD) technique on the \lcs, whereby deviations
from the median were cross-correlated with a number of ``external
parameters'', such as the $X$ and $Y$ sub-pixel position, FWHM,
hour-angle, and zenith distance. We have also applied the Trend
Filtering Algorithm \citep[TFA;][]{kovacs05} along with the Box Least
Squares \citep[BLS;][]{kovacs02} transit-search algorithm in our
analysis. For field G196 we gathered $\sim$3,750 (HAT-7) plus $\sim$890
(HAT-9) data-points at 5.5\,min cadence between 2005 June 8 and 2005
December 5 (UT). In the \lc\ of star GSC~02634-01087 we detected a
$\sim$13\,mmag transit with a 2.7881\,d period, signal-to-noise ratio
of $12$ in the BLS frequency spectrum, and dip-significance of $18$
\citep{kovacs05b}. The top panel of \figr{lc} shows the unbinned \lc\
with all $\sim$4640 data points, folded with the period that we derived
subsequently, based on high precision follow-up, as described later in
\secr{photfol}.

\subsection{Early spectroscopy follow-up}
\label{sec:ds}

Initial follow-up observations were made with the CfA Digital
Speedometer \citep[DS;][]{latham92} in order to characterize the host
star and to reject obvious astrophysical false-positive scenarios that
mimic planetary transits. The four radial velocity (RV) measurements
obtained over an interval of 33 days showed an rms residual of
0.41\,\kms, consistent with no detectable RV variation. Atmospheric
parameters for the star (effective temperature $T_{\rm eff}$, surface
gravity $\log g$, metallicity [Fe/H], and projected rotational velocity
$v \sin i$) were derived as described by \cite{torres02}. The first
three quantities are strongly correlated and difficult to determine
simultaneously. For example, the unconstrained value $\log g = 4.0 \pm
0.2$ we obtained is somewhat lower than derived from our stellar
evolution modeling in \secr{stelpar}, which is $\log g = 4.37$.
Consequently, in a second iteration we held $\log g$ fixed at this
value and redetermined the other quantities, obtaining $\teff = 5960
\pm 100$~K, [Fe/H] $= +0.24 \pm 0.15$, and $\vsini = 2.6 \pm 1.5$~\kms.
These correspond to a slowly-rotating early G main sequence star.

\subsection{High-precision spectroscopy follow-up}
\label{sec:hires}

\notetoeditor{This is the intended place of \tabr{rv}.}
\begin{deluxetable}{lrrr}
\tabletypesize{\scriptsize}
\tablewidth{0pt}
\tablecaption{
	\label{tab:rv}
	Radial Velocities for \hatcur.
}
\tablewidth{0pt}
\tablehead{
	\colhead{BJD $- 2,\!400,\!000$} &
	\colhead{RV\tablenotemark{a}} &
	\colhead{$\sigma_{RV}$} &
	\colhead{BS\tablenotemark{b}} \\
	\colhead{(days)} &
	\colhead{(\ms)} &
	\colhead{(\ms)} &
	\colhead{(\ms)} \\
}
\startdata
54227.5199 &   7721.4 &         12.2	&  22.5	\\
54228.5949 &   7457.4 &         22.3	&  18.2	\\
54229.6098 &   7710.4 &         17.2	&  -7.0	\\
54230.4900 &   7603.8 &         22.4	& -20.0	\\
54231.6088 &   7521.3 &         14.3	&  15.2	\\
54233.6057 &   7579.4 &		  15.1	& -31.0	\\
54234.5210 &   7510.5 &         21.3	& -54.8	\\
54255.5171 &   7680.5 &          9.8	&   4.2
\enddata
\tablenotetext{a}{The RVs include the barycentric correction.}
\tablenotetext{b}{Bisector spans.}
\label{SOPHIE}
\end{deluxetable}

High-resolution spectroscopic follow-up was carried out at the Haute
Provence Observatory (OHP) 1.93-m telescope, with the SOPHIE
spectrograph \citep{bouchy06}. SOPHIE is a multi-order echelle
spectrograph fed through two fibers, one of which is used for starlight
and the other for sky background or a wavelength calibration lamp. The
instrument is entirely computer-controlled and a standard data
reduction pipeline automatically processes the data upon CCD readout. 
RVs are calculated by numerical cross-correlation with a high
resolution observed spectral template of a G2 star. Similar
spectroscopic follow-up with SOPHIE has already resulted in the
confirmation of two TEPs: WASP-1b and WASP-2b \citep{cameron07}. \hatcur\
was observed with SOPHIE in the high-efficiency mode (R $\sim 39000$)
during our May 2--13, 2007 observing run, with an additional
measurement taken on June 4. Depending on observing conditions,
exposure times were in the range of 15 to 35 minutes, resulting in
signal to noise ratios of 20--55 per pixel at $\lambda = 5500$\,\AA.
Using the empirical relation of Cameron et al.\ (2007) we estimated the
RV photon-noise uncertainties to be 10--25\ms. We present here 8 radial
velocity measurements taken when the planet was out of transit, listed
in Table~\ref{SOPHIE}.  Measurements taken during transit, revealing
the Rossiter-McLaughlin effect \citep[e.g.][]{winn05}, will be
presented in a separate paper.

\subsection{Photometry follow-up}
\label{sec:photfol}

In order to better characterize the transit parameters and also to
derive a better ephemeris, we performed follow-up photometric observations with 1-m
class telescopes. A partial transit of \hatcurb\ was observed using the
KeplerCam detector on the \flwof\ telescope \citep[see][]{holman07} on
UT 2007 May 18. We refer to this event as having transit number $N_{tr}
= -1$. Three days later a full transit, $N_{\mathrm{tr}}=0$, was
observed with the same instrument. The two Sloan $z$-band light curves
are shown in the middle panel of \figr{lc}. We also gathered data for
four subsequent full transit events, $N_{\mathrm{tr}}=2$,
$N_{\mathrm{tr}}=11$, $N_{\mathrm{tr}}=16$ and $N_{\mathrm{tr}}=21$,
using the \wom\ telescope in the Cousins $R$ band (lower panel of
\figr{lc}). Data were reduced in a similar manner to the HATNet data,
using aperture photometry and an ensemble of $\sim$300 comparison stars
in the field.  Since the follow-up observations span 22 transit cycles
($\sim$2 month time-span), we were able to obtain an accurate
ephemeris. An analytic model was fit to these data, as described below
in \secr{planpar}, and yielded a period of \hatcurP\,d and reference
epoch of mid-transit $T_{\mathrm{c}}=\hatcurT$\,d (HJD).  The length of
the transit as determined from this joint fit is \hatcurdur\,d (2
hours, 55 minutes), the length of ingress is \hatcuringdur\,d (20.9
minutes), and the central transit depth is 0.0136\,mag.

\section{Stellar parameters}
\label{sec:stelpar}

\notetoeditor{This is the intended place of \tabr{stelpar}}
\begin{deluxetable}{lll}
\tabletypesize{\scriptsize}
\tablecaption{
	Summary of stellar parameters for \hatcur.
\label{tab:stelpar}}
\tablehead{
	\colhead{Parameter} &
	\colhead{Value} &
	\colhead{Source}
}
\startdata
	\teff (K)		&	$5960\pm100$	& DS \\
	\vsini (\kms)		&	$2.6\pm 1.5$	& DS \\
	\logg 			&	$4.368\pm0.028$	& Yonsei-Yale \\
	\feh (dex)		&	$+0.24\pm0.15$	& Yonsei-Yale \\
	Distance (pc)\tablenotemark{a}		&	$340\pm30$	& Yonsei-Yale \\
	Mass (\msun)		&	$1.160\pm0.062$	& Yonsei-Yale \\
	Radius (\rsun)		&	$1.167\pm0.049$	& Yonsei-Yale \\
	$\log (\lstar /\lsun)$	&	$0.187\pm0.064$	& Yonsei-Yale \\
	$M_V$			&	$4.32\pm0.18$	& Yonsei-Yale \\
	Age (Gyr)		&	$2.6\pm1.8$		& Yonsei-Yale
\enddata
\tablenotetext{a}{Assuming no extinction due to the proximity of the
star.}
\end{deluxetable}

The mass (\mpl) and radius (\rpl)  of a transiting planet scale with
those of the parent star. In order to determine the stellar properties
needed to place \mpl\ and \rpl\ on an absolute scale, we made use of
stellar evolution models along with the observational constraints from
spectroscopy. Because of its relative faintness, the host star does not
have a parallax measurement from {\it Hipparcos\/}, and thus a direct
estimate of the absolute magnitude is not available for use as a
constraint. An alternative approach is to use the surface gravity of
the star, which is a sensitive measure of the evolutionary state of the
star and therefore has a very strong influence on the radius. However,
$\logg$ is a notoriously difficult quantity to measure
spectroscopically and is often strongly correlated with other
spectroscopic parameters (see \secr{ds}).  It has been pointed out by
\citet{sozzetti07} that the normalized separation of the planet,
$a/\rstar$, can provide a much better constraint for stellar parameter
determination than \logg. The $a/\rstar$ quantity can be determined
directly from the photometric observations with no additional
assumptions (other than limb-darkening, which is a second-order
effect), and it is related to the density of the central star. As
discussed later in \secr{planpar}, an analytic fit to the light curve
yields $a/\rstar = 7.50\pm0.19$.

This value, along with $T_{\rm eff}$ and [Fe/H] from \secr{planpar},
was compared with the Yonsei-Yale stellar evolution models of
\cite{yi01} following \citet{sozzetti07}. As described earlier, the
initial temperature and metallicity from our DS spectroscopy were
subsequently improved by applying the $\log g$ constraint from the
models, and the isochrone comparison was repeated. This resulted in
final values for the stellar mass and radius of $\mstar = 1.160 \pm
0.062\,\msun$ and $\rstar =1.167 \pm 0.049\,\rsun$, and an estimated
age of $2.6 \pm 1.8$\,Gyr. We summarize these and other properties in
\tabr{stelpar}.

\section{Spectroscopic orbital solution}
\label{sec:rv}

\notetoeditor{This is the intended place of \figr{rv}. We would like to 
typeset it as a single column figure.}
\ifthenelse{\boolean{emulateapj}}{\begin{figure}[t]}{\begin{figure}[t]}
\ifpdf
\plotone{img/plotbisrv.pdf}
\else
\plotone{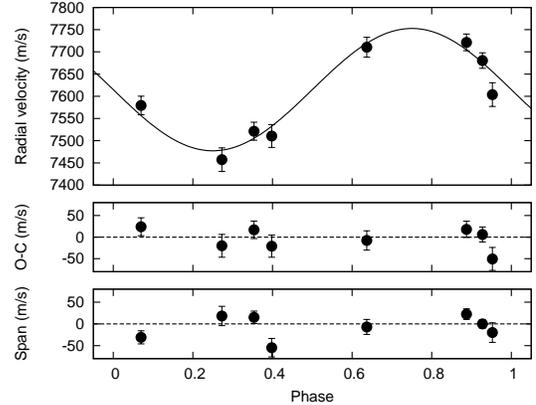}
\fi
\caption{
	The top panel shows the RV measurements phased with the period of
	$P=2.788491\,d$. The zero-point in phase corresponds to the epoch
	of mid-transit. Overlaid is the best fit, assuming 14.4\,\ms\
	stellar jitter. The middle panel shows the residuals from the fit. 
	The bottom panel displays the line bisector spans on the same scale
	as the upper panel. No variation in the line bisectors is seen
	concomitant with that in the RVs, essentially confirming the
	planetary nature of the transiting object.
\label{fig:rv}}
\ifthenelse{\boolean{emulateapj}}{\end{figure}}{\end{figure}}

Our 8 RV measurements from SOPHIE were fitted with a Keplerian orbit
model solving for the velocity semi-amplitude $K$ and the
center-of-mass velocity $\gamma$, holding the period and transit epoch
fixed at the well-determined values from photometry. The eccentricity
was initially set to zero. The resulting rms residual of
$\sim$23.7~m~s$^{-1}$ is somewhat larger than expected from the
internal errors, and we find that a reduced $\chi^2$ value of unity
necessitates the addition of uncorrelated noise of 14.4~m~s$^{-1}$ in
quadrature, which we attribute to ``stellar jitter''. This level of
jitter is consistent with the predictions of \citet{saar98} for a
projected rotational velocity such as what we measure for the parent
star.
The final fit, with the internal errors increased as described above,
yields $K=138\pm14$\ms\ and $\gamma=7613.8\pm9.1$\ms. The observations
and fitted RV curve are displayed in the top panel of \figr{rv}, with
the residuals shown in the middle panel.

As a test we allowed for the possibility of an eccentric orbit and
solved for the two additional quantities $e\cos\omega$ and
$e\sin\omega$, but the results were insignificantly different from
zero.

\section{Excluding blend scenarios}
\label{sec:blend}

We have tested the reality of the velocity variations by carefully
examining the spectral line bisectors of the star using our OHP data.
If the velocity changes measured are due only to distortions in the
line profiles arising from contamination of the spectrum by the
presence of a binary with a period of 2.79 days, we would expect the
bisector spans (which measure line asymmetry) to vary with this period
and with an amplitude similar to the velocities \citep[see,
e.g.,][]{queloz01,torres05}. As shown in the lower panel of \figr{rv},
the changes in the bisector spans are of the same order as the residual
RV variations, and much smaller than the radial velocity semi-amplitude
itself.  This analysis shows that the orbiting body is a planet and
rules out a possible blend scenario.

\section{Planetary parameters}
\label{sec:planpar}

\notetoeditor{This is the intended place of \tabr{orb}}
\begin{deluxetable}{lr}
\tablecaption{
	\label{tab:orb}
	Orbital fit and planetary parameters for the \hatcurb\ system.
}
\tablehead{
	\colhead{Parameter} &
	\colhead{Value}
} 
\startdata
Period (d)\tablenotemark{a}					& 	\hatcurP		\\
$T_{\mathrm{mid}}$ (HJD)\tablenotemark{a}	& 	\hatcurT		\\
Transit duration (day)						&	\hatcurdur		\\
Ingress duration (day)						&	\hatcuringdur	\\
\hline
Stellar jitter (\ms)\tablenotemark{b}		&	14.4			\\
$\gamma$ (\ms)\tablenotemark{c}				&	$7613.8\pm9.1$	\\
$K$ (\ms)									&	$138\pm14$		\\
$e$ \tablenotemark{a}						&	$0$				\\
\hline
$a_{\rm rel}/R_\star$						&	\hatcurar	\\
$\rpl/\rstar$								&	\hatcurrprstar	\\
$a_{\rm rel}$ (AU)							&	\hatcurarel	\\
\ipl (deg)									&	\hatcuri	\\
\mpl (\mjup)								&	\hatcurm	\\
\rpl (\rjup)								&	\hatcurr	\\
\rhopl (\gcmc)								&	\hatcurrho	\\
\gpl ($m\,s^{-2}$)\tablenotemark{d}			&	\hatcurg	\\
\hline
$\Delta T_{c,0}$, $N_{tr}=0$ ($10^{-5}d$)	&	$6\pm27$	\\
$\Delta T_{c,0}$, $N_{tr}=2$ ($10^{-5}d$)	&	$79\pm58$	\\
$\Delta T_{c,0}$, $N_{tr}=11$ ($10^{-5}d$)	&	$6\pm62$	\\
$\Delta T_{c,0}$, $N_{tr}=16$ ($10^{-5}d$)	&	$-112\pm84$	\\
$\Delta T_{c,0}$, $N_{tr}=21$ ($10^{-5}d$)	&	$11\pm57$
\enddata
\tablenotetext{a}{Fixed in the orbital fit.}
\tablenotetext{b}{Adopted (see text).}
\tablenotetext{c}{The $\gamma$ velocity is on an absolute reference frame.}
\tablenotetext{d}{Based on only directly observable quantities, 
	see \citet{southworth07}.}
\end{deluxetable}

The light-curve parameters of \hatcurb\ were determined from a joint
fit based on the 6 distinct transit events, observed with the \flwof\
and \wom\ telescopes. A circular orbit was assumed, based on our
analysis above. We adopted a quadratic limb-darkening law for the star,
and took the appropriate coefficients from \citet{claret04}, for both
the Sloan $z$ and Cousins $R$ bands. The drop in flux in the \lcs\ was
modeled with the formalism of \citet{mandel02}, using the equations for
the general case (i.e.~{\em not} the small planet approximation).  The
adjusted parameters in the fit were
i) the mid-transit times of the first full transit ($N_{tr} = 0$,
$T_{c0}$), and the last full transit ($N_{tr} = 21$, $T_{c21}$),
(this is equivalent to fitting for an epoch $E$ 
and a period $P$), 
ii) the relative planetary radius, $p\equiv R_p/\rstar$; 
iii) the square of the normalized impact parameter, $b^2$; 
iv) the quantity $\zeta/\rstar \equiv a/\rstar(2\pi/P)/\sqrt{1-b^2}$.
From simple geometric considerations $\zeta/R_\star$ and $b^2$ have an
uncorrelated {\em a posteriori} probability distribution in parameter
space. This amounts to an orthogonalization of the fitted parameters,
similar (albeit simpler) to the one employed by \cite{burke07} for the
case of XO-2b.

We used the Markov Chain Monte Carlo algorithm \citep[see,
e.g.][]{holman07} to derive the best fit parameters.  Uncertainties
were estimated using synthetic data sets, by added Gaussian noise to
the fitted curve at the dates of our observations, and re-solving the
\lc\ analytic fit. The magnitude of the noise was taken from the white-
and red-noise estimations based on the real residuals. This process was
repeated $1.5\cdot10^5$ times, yielding a good representation of the
{\em a posteriori} distribution of the best-fit parameter values. We
found this method of error estimation to be robust, since it is not
sensitive to the number of out-of-transit (OOT) points.

The result for the radius ratio is $\rpl/\rstar=0.1106\pm0.0006$, and
the normalized separation is $a/\rstar = \hatcurar$. We found that the
\emph{a posteriori} distribution of $b^2$ is consistent with a
symmetric Gaussian distribution, and yields $b^2=0.181\pm0.040$,
therefore the orbit is inclined.
From the inclination, the mass of the star (\tabr{stelpar}), and the
orbital parameters (\secr{rv}), the other planetary parameters (such as
mass, radius) are derived in a straightforward way, and are summarized
in \tabr{orb}.
We note that $a/\rstar$, as derived from the \lc\ fit, is an important
constraint in the stellar parameter determination (\secr{stelpar}),
which in turn defines the limb-darkening coefficients that are used in
the \lc\ analytic fit.  Thus, after the initial analytic fit to the
\lc\ and the stellar parameter determination, we performed another
iteration in the \lc\ fit. We found that the change in parameters was
imperceptible.

The possibility of transit time variations (TTVs) was checked by
fitting the center of the transit of the five full transit events
independently. We found no sign of TTV, as the transit times differ by
less than 1-$\sigma$ from the expected values (listed in \tabr{orb}).

\section{Conclusions}
\label{sec:disc}

\hatcurb\ is an ordinary hot Jupiter (P = 2.788 days) with slightly
inflated radius (\rpl = 1.26\rjup) for its mass of 1.06\mjup, orbiting
a slightly metal rich solar-like star. The $\sim$20\% radius inflation
is what current models predict for a planet with equilibrium
temperature of $\sim$1500K \citep{burrows07,fortney07}.

\hatcurb\ is more massive than any of the known TEPs with similar
period ($2.5\lesssim P \lesssim3$\,d), such as XO-2b, WASP-1, HAT-P-3b,
TRES-1, and HAT-P-4b, with the exception of TRES-2. The latter is
fairly similar in mass, radius, orbital period, and stellar effective
temperature.

However, \hatcurb\ is interesting in that it falls between Class I and
II, as defined by the Safronov number and $T_{eq}$ of the planet
\citep{hansen07}. \hatcurb\ has a Safronov number of $0.059\pm0.005$ ,
while Class I is defined as $0.070\pm0.01$, especially at $T_{eq}\sim
1500$K. It seems that the additional discovery and characterization of
transiting planets of Jupiter and higher masses would be very helpful
in order to understand these new correlations and their reality.

\acknowledgments

Operation of the HATNet project is funded in part by NASA grant
NNG04GN74G.
Work by G\'AB was supported by NASA through Hubble
Fellowship Grant HST-HF-01170.01-A.
GK wishes to thank support from Hungarian Scientific Research
Foundation (OTKA) grant K-60750.
We acknowledge partial support from the Kepler Mission under NASA
Cooperative Agreement NCC2-1390 (DWL, PI).
A.P.~would like to thank the hospitality of the CfA, where this work 
has been carried out. A.P.~was also supporeted by the
Doctoral School of the E\"otv\"os University.
GT acknowledges partial support from NASA Origins grant NNG04LG89G.
TM thanks the Israel Science Foundation for a support through grant
no.~03/233.
We owe special thanks to the directors and staff of FLWO and SMA for
supporting the operation of HATNet.  We thank the OHP and SOPHIE team
for their help in carrying the observations that have been funded by
OPTICON.




\end{document}